\documentclass[12pt]{iopart}

\usepackage{iopams}  
\usepackage{graphicx}

\begin{document}
\title{A weak-value interpretation of the Schwinger mechanism of massless/massive pair productions}

\author{Kazuhiro Yokota and Nobuyuki Imoto}
\address{Department of Materials Engineering Science,
Graduate School of Engineering Science, Osaka University,
Toyonaka, Osaka 560-8531, Japan}
\ead{yokota@qi.mp.es.osaka-u.ac.jp}
\date{\today}
\begin{abstract}
According to the Schwinger mechanism, a uniform electric field brings about pair productions in vacuum; the relationship between the production rate and the electric field is different, depending on the dimension of the system.
In this paper, we make an offer of another model for the pair productions, in which weak values are incorporated:
energy fluctuations trigger the pair production, and a weak value appears as the velocity of a particle there.
Although our model is only available for the approximation of the pair production rates, the weak value reveals a new aspect of the pair production.
Especially, within the first order, our estimation approximately agrees with the exponential decreasing rate of the Landau-Zener tunneling through the mass energy gap.
In other words, such tunneling can be associated with energy fluctuations via the weak value, when the tunneling gap can be regarded as so small due to the high electric field.
\end{abstract}

\maketitle

\section{Introduction}
\label{sec:introduction}
Weak value, introduced as a result of weak measurements \cite{W1}, has attracted attention in quantum physics recently \cite{W2}.
While the application of the amplification effect (AAV effect) has been discussed actively, the weak value gives us a new approach to the fundamental issues.
At the same time, it has been challenging to shed light on the meaning of a weak value, since it may take an anomalous value lying outside the range of eigenvalue spectra \cite{W3, W4, W5}.
In particular, it has been of great importance to reveal whether a weak value is worthy of a physical value by itself, namely, without the context of measurements; weak values were found to be useful for describing several quantum phenomena \cite{W_ph1, W_ph2, W_ph3, W_ph4}.
Furthermore, it should be verified that a weak value works in quantum physics not only qualitatively but also quantitatively, by which the weak value must attain the position of a value of a physical quantity.
In fact, we gave indication of the appearance of a weak value in the Dirac equation;
with the weak value, we could describe the pair production via a supercritical step potential \cite{step} and the supply of an electric carrier in graphene \cite{gra_wv}.
Importantly, we also succeeded in estimating the pair production rate and the electric current respectively in these previous works: we can cope with the quantitative problems by means of the weak values.

In this paper, along the lines of these discussions, we deal with the Schwinger mechanism in the view of weak values.
The Schwinger mechanism represents pair productions in vacuum triggered by a uniform electric field $\varepsilon$, the rate of which in $l+1$ dimension ($l=1,2$) is given by
\begin{eqnarray}
\left(\frac{dn}{dt}\right)_{l+1} = \frac{(q \varepsilon)^{(l+1)/2}}{(2\pi)^l\hbar^{(l+1)/2}c^{(l-1)/2}}{\rm exp}\left(-\frac{\pi m^2c^3}{q\varepsilon \hbar}\right) \ \ \ \ \ \ (l = 1,2), \label{eq:i_rate}
\end{eqnarray}
where the particle has the mass $m$ and the charge $q$; $c$ is the velocity of light \cite{Schwinger1, Schwinger2, Schwinger3}.
Because of the two degrees of freedom by the 1/2 spin, which makes double counts, the rate in 3+1 dimension is similarly given by
\begin{eqnarray}
\left(\frac{dn}{dt}\right)_{3+1} = \frac{(q \varepsilon)^2}{4\pi^3\hbar^2c}{\rm exp}\left(-\frac{\pi m^2c^3}{q\varepsilon \hbar}\right).  \label{eq:i_rate3}
\end{eqnarray}
The exponential decrease in the rates comes from the Landau-Zener tunneling through the energy gap between $-mc^2$ and $mc^2$.
In \cite{gra_wv}, using weak values, we estimated the production rate for the massless ($m=0$) pairs in 2+1 dimension, when we discussed the electric current in graphene.
In similar fashion, it should be also possible to derive the rates for massless pairs in the other dimensions by using weak values, which is our first goal.
For the massive pairs, when the electric field is strong enough to satisfy $m^2c^3\ll q\varepsilon\hbar$, in all the dimension, the decreasing rate by the mass can be obtained as follows,
\begin{eqnarray}
\left(\Delta\frac{dn}{dt}\right)_{l+1} &\equiv & \frac{\left(\frac{dn}{dt}\right)^{m=0}_{l+1}-\left(\frac{dn}{dt}\right)_{l+1}}{\left(\frac{dn}{dt}\right)^{m=0}_{l+1}} \ \ \ \ \ \ (l=1,2,3) \\ 
&=& 1-{\rm exp}\left(-\frac{\pi m^2c^3}{q\varepsilon \hbar}\right) \ \sim \ \frac{\pi m^2c^3}{q\varepsilon \hbar} \ \ \ \ \ \ \left(\frac{m^2c^3}{q\varepsilon \hbar}\ll 1\right), \label{eq:i_rate_m}
\end{eqnarray}
where $\left(\frac{dn}{dt}\right)^{m=0}_{l+1}$ denotes the production rate for massless pairs.
Our second goal is to obtain this decreasing rate in each dimension.
In discussing such quantitative problem, the weak value offers a new qualitative aspect of the pair production, especially, the Landau-Zener tunneling in the Schwinger mechanism.
The singularity of the massless, 1+1 dimensional case is also clarified in estimating the pair production rate, because the pair production takes place in the different manner from the other cases.
In addition, we can also make the deeper understanding of how the weak values derive the results in the consistent way, when we consider the massive cases:
we can clarify the difference between transition and transmission, which was not shown in the previous paper \cite{gra_wv}.

Hereafter the paper is organized as follows.
In section \ref{sec:2+1}, we review our previous result of the massless Dirac particles in 2+1 dimension and generalise it for the massive ones.
This procedure can be easily applied to the 3+1 dimensional case in section \ref{sec:3+1}.
In section \ref{sec:1+1}, we deal with the 1+1 dimensional case, in which we need to alter our procedure a little bit: we need a slightly artful treatment for the massless pairs because of the singularity as we mentioned above.
We conclude our result in section \ref{sec:conclusion}.

\section{The 2+1 dimensional case}
\label{sec:2+1}
In this section, we first review our previous result for the massless, 2+1 dimensional case \cite{gra_wv}, and then the theory is developed into the massive, 2+1 dimensional case.
\subsection{The massless pairs}
Using the Pauli matrices $\hat{\sigma}_k$, the Hamiltonian of a massless Dirac particle in 2+1 dimension can be represented as follows,
\begin{eqnarray}
\hat{H}_{2+1}^{m=0}=c(\hat{\sigma}_x\hat{p}_x+\hat{\sigma}_y\hat{p}_y)-q\varepsilon x,
\end{eqnarray}
where we have assumed the uniform electric field $\varepsilon$ in $x$ direction.
An eigenstate of the free Hamiltonian with an energy $E$ and a momentum $\vec{p}=(p_x, p_y)$ can be described by the chirality and the space part as $|E,p_x,p_y\rangle\psi_{p_x,p_y}(x,y)$, where the chirality $|E,p_x,p_y\rangle$ is independent of $x$ and $y$:
the chirality of a negative (positive) energy $-E$ ($E$) is given by
\begin{eqnarray}
|\pm E,p_x,p_y\rangle = \frac{1}{\sqrt{2}}
\left[
 \begin{array}{c}
 e^{-i\theta /2}\\
 \pm e^{i\theta /2}
 \end{array}
\right]
= \frac{1}{\sqrt{2E}}
\left[
 \begin{array}{c}
 (p_x-ip_y)c\\
 \pm E
 \end{array}
\right],
\end{eqnarray}
up to the global phase, where $\theta={\rm Arctan}(p_y/p_x)$ and $E^2=p^2c^2=(p_x^2+p_y^2)c^2$.
Since an energy eigenstate can be specified by the chirality, an energy transition can be represented by a pre-postselection on the chirality;
in particular, a transition from a negative energy $-E<0$ to a positive one $E'>0$ corresponds to a particle-antiparticle pair production.
Considering the direction of the electric field, we discuss the transition from ($-E,-p_x,p_y$) to ($E',p_x',p_y$) with $p_x, p_x'$ $>0$, by which the directions of the velocities before and after the transition are the same as in $+x$ direction.
We have also taken account of the preservation of the momentum in $y$ direction.
When the time $t$ is small enough, the pre-postselection on the chirality brings about the time evolution on the space part $\psi_{-p_x,p_y}(x,y)$ as follows
\footnote{
There are typos in \cite{gra_wv}.
`preselection $|E\rangle$' should be corrected as `preselection $|-E\rangle$.'
In addition to the energy, the momentum is also involved in the chirality.
In this paper, we clarify it in the notation of the chirality as $|E,p_x,p_y\rangle$ to avoid confusion.
},
\begin{eqnarray}
& &\langle E',p_x',p_y| e^{-\frac{i}{\hbar}\hat{H}_{2+1}^{m=0}t}|-E,-p_x,p_y\rangle \psi_{-p_x,p_y}(x,y)  \label{eq:t_ev2} \\
&\sim& \langle E',p_x',p_y|-E,-p_x,p_y\rangle e^{-\frac{i}{\hbar}c\langle\hat{\sigma}_x\rangle _{\bf w}\hat{p}_xt}     \nonumber \\
& & \ \ \ \ \ \ \ \ \ \ \ \ \ \ \ \ \ \ \ \ \ \ \ \ \ \ \ \ \ \ \ \ \ e^{-\frac{i}{\hbar}c\langle\hat{\sigma}_y\rangle_{\bf w}\hat{p}_yt}e^{\frac{i}{\hbar}q\varepsilon x t}\psi_{-p_x,p_y}(x,y)  \ \ \ (t \ \sim \ 0) \label{eq:t_ev_app2} \\
&\sim& \langle E',p_x',p_y|-E,-p_x,p_y\rangle \psi_{-p_x+q\varepsilon t,p_y}(x-c\langle\hat{\sigma}_x\rangle_{\bf w}t,y-c\langle\hat{\sigma}_y\rangle_{\bf w} t) \nonumber \\
& & \ \ \ \ \ \ \ \ \ \ \ \ \ \ \ \ \ \ \ \ \ \ \ \ \ \ \ \ \ \ \ \ \ \ \ \ \ \ \ \ \ \ \ \ \ \ \ \ \ \ \ \ \ \ \ \ \ \ \ \ \ \ \ \ \ \ \ \ \ \ \ \ \ \ (t \ \sim \ 0),  \label{eq:WVshift2}
\end{eqnarray}
where $\langle \hat{\sigma}_k\rangle_{\bf w}$ $(k=x,y)$ is a weak value defined by,
\begin{eqnarray}
\langle\hat{\sigma}_k\rangle_{\bf w} = \frac{\langle E',p_x',p_y|\hat{\sigma}_k|-E,-p_x,p_y\rangle}{\langle E',p_x',p_y|-E,-p_x,p_y\rangle}.   \label{eq:WV}
\end{eqnarray}
$c\langle\hat{\sigma}_k\rangle_{\bf w}$ corresponds to the group velocity in $k$ direction and gives the current driven by the transition.
Since the electric field is in $x$ direction, $\langle\hat{\sigma}_y\rangle_{\bf w}$ should be zero, from which we can find
\begin{eqnarray}
p_x'=p_x \ {\rm and} \ E'=E.  \label{eq:selective}
\end{eqnarray}
That is to say, only the transition from ($-E,-p_x,p_y$) to ($E,p_x,p_y$) is allowed.
In this case, the weak value of $\hat{\sigma}_x$ shows
\begin{eqnarray}
\langle \hat{\sigma}_x\rangle_{\bf w} = \frac{\sqrt{p_x^2+p_y^2}}{p_x}=\frac{E}{p_xc},  \label{eq:WV2}
\end{eqnarray}
where we have made use of $E^2=p^2c^2=(p_x^2+p_y^2)c^2$.
As shown in \cite{gra_wv}, the velocity given by this weak value is the requisite velocity to yield the changing of both the energy $\Delta E=2E$ and the momentum $\Delta p_x=2p_x$: the electric field should perform the work $q\varepsilon \Delta x =\Delta E$ and the impulse $q\varepsilon \Delta t=\Delta p_x$ with the distance $\Delta x$ and the time $\Delta t$, by which the average velocity of the particle can be defined as $\Delta x/\Delta t=\Delta E/\Delta p_x$.
In fact, this average velocity agrees with $c\langle\hat{\sigma}_x\rangle_{\bf w}$.

Although $c\langle\hat{\sigma}_x\rangle_{\bf w}$ is larger than $c$, the velocities in $x$ direction just before and after the transition are both given by $p_xc^2/E<c$.
With the transition probability $T(p)$, they should satisfy,
\begin{eqnarray}
T(p)c\langle\hat{\sigma}_x\rangle_{\bf w}=\frac{p_xc^2}{E} = \left(\frac{E}{p}\frac{p_x}{\sqrt{p_x^2+p_y^2}}\right),   \label{eq:cons_WV2}
\end{eqnarray}
by which the weak value can hold the consistency: the average velocities should be equivalent.
Note that, without a transition, a velocity never comes into being.
In other words, zero velocity with the probability $1-T(p)$ is included in l.h.s. of equation (\ref{eq:cons_WV2}).
As a result, the transition probability is given by
\begin{eqnarray}
T(p) = \frac{p_x^2}{p_x^2+p_y^2}.   \label{eq:prob_2}
\end{eqnarray}

In \cite{gra_wv}, we assumed that the transition can be triggered by virtual particles (virtual transitions) allowed by the uncertainty relations, within which the time is so small that the the higher terms of $O(t^k)$ ($k\ge 2$) are smaller than the first one of $O(t)$ in the time evolution of equation (\ref{eq:t_ev2}); using the weak values, we can approximately deal with the quantitative problem, namely, the pair production rate, although the time does not always satisfy $t\sim 0$.
According to the energy-time uncertainty relation, an energy fluctuation corresponding to $\Delta E$ can be allowed during the time $\delta t$ which satisfies,
\begin{eqnarray}
\delta t = \frac{\hbar}{\Delta E}.
\end{eqnarray}
A virtual transition caused by this fluctuation can be made real via the weak value, which provides the requisite velocity for changing both the energy and the momentum as mentioned above.
A quantum state can participate in a pair production, if the needed time for the transition $\Delta t$ satisfies $\Delta t \le \delta t$, from which we can obtain,
\begin{eqnarray}
p_x^2(p_x^2+p_y^2)\le \frac{q^2\varepsilon^2\hbar^2}{16c^2}.   \label{eq:qs_20}
\end{eqnarray}
Then, we can estimate the pair production rate as follows, 
\begin{eqnarray}
\left(\frac{dn}{dt}\right)_{2+1}^{m=0} &=& \frac{1}{(2\pi\hbar)^2}\int \!\!\! \int_{(\ref{eq:qs_20}) \ {\rm and} \ p_x>0}dp_x dp_y\frac{T(p)}{\Delta t}  \\
&=& \frac{q\varepsilon}{2(2\pi\hbar)^2}\int \!\!\! \int_{(\ref{eq:qs_20}) \ {\rm and} \ p_x>0}dp_xdp_y\frac{p_x}{p_x^2+p_y^2}     \label{eq:rate2}   \\
&=&\frac{(q\varepsilon)^{3/2}}{4\pi^2\hbar ^{3/2}c^{1/2}}\frac{1}{2}\int_{0}^{1} ds \ {\rm Arctan}  \left( \sqrt{\frac{1}{s^4}-1} \right)  \\
&=&\frac{B(1/2,3/4)}{4}\frac{(q\varepsilon)^{3/2}}{4\pi^2\hbar ^{3/2}c^{1/2}} \ \sim \ 0.60\frac{(q\varepsilon)^{3/2}}{4\pi ^2\hbar ^{3/2}c^{1/2}},    \label{eq:rate2_0}
\end{eqnarray}
where $B(m,n)$ represents the beta function.
Note that the production rate in $(p_x, p_y)\sim(p_x+dp_x, p_y+dp_y)$ is given by $\frac{1}{(2\pi\hbar)^2}dp_xdp_x \frac{T(p)}{\Delta t}$, because $\frac{T(p)}{\Delta t}$ represents the rate of a quantum state contributing to the pair productions per unit time.
This result approximately agrees with the pair production rate in equation (\ref{eq:i_rate}) with $l=2$ and $m=0$.

\subsection{The massive pairs}
The procedure reviewed above can be applied to the massive Dirac particle, which is subjected to the Hamiltonian given by
\begin{eqnarray}
\hat{H}_{2+1}=c(\hat{\sigma}_x\hat{p}_x+\hat{\sigma}_y\hat{p}_y)+mc^2\hat{\sigma}_z-q\varepsilon x.
\end{eqnarray}
As in the massless particle, an eigenstate of the free Hamiltonian with a negative (positive) energy $-E$ ($E$) and a momentum $\vec{p}=(p_x, p_y)$ can be assigned with the chirality as follows,
\begin{eqnarray}
|\pm E,p_x,p_y\rangle = \frac{1}{\sqrt{2E(E\mp mc^2)}}
\left[
 \begin{array}{c}
 (p_x-ip_y)c\\
 \pm E-mc^2
 \end{array}
\right].
\end{eqnarray}
Then, considering the transition from a negative energy state ($-E,-p_x,p_y$) to a positive one ($E',p'_x,p_y$), the time evolution of the space part $\psi_{-p_x,p_y}(x,y)$ can be given by
\begin{eqnarray}
& &\langle E',p_x',p_y| e^{-\frac{i}{\hbar}\hat{H}_{2+1}t}|-E,-p_x,p_y\rangle \psi_{-p_x,p_y}(x,y)               \\
&\sim& \langle E',p_x',p_y|-E,-p_x,p_y\rangle e^{-\frac{i}{\hbar}c\langle\hat{\sigma}_x\rangle _{\bf w}\hat{p}_xt} \nonumber \\
& & \ \ \ \ \ \ \ \ \ \ \ \ \ \ \ \ \ \ \ e^{-\frac{i}{\hbar}c\langle\hat{\sigma}_y\rangle_{\bf w}\hat{p}_yt}
e^{-\frac{i}{\hbar}mc^2\langle\hat{\sigma}_z\rangle_{\bf w}t}
e^{\frac{i}{\hbar}q\varepsilon x t}\psi_{-p_x,p_y}(x,y)  \ \ \ (t \ \sim \ 0)      \label{eq:t_ev_app2m}   \\
&\sim& \langle E',p_x',p_y|-E,-p_x,p_y\rangle e^{-\frac{i}{\hbar}mc^2\langle\hat{\sigma}_z\rangle_{\bf w}t}  \nonumber \\
& & \ \ \ \ \ \ \ \ \ \ \ \ \ \ \ \ \ \ \ \psi_{-p_x+q\varepsilon t,p_y}(x-c\langle\hat{\sigma}_x\rangle_{\bf w}t,y-c\langle\hat{\sigma}_y\rangle_{\bf w} t)  \ \ \ (t \ \sim \ 0).
\end{eqnarray}
This approximation is also reasonable as long as the time evolution stems from a virtual transition allowed by the uncertainty relation as mentioned above.
If equation (\ref{eq:t_ev_app2m}) gives just the shift in $x$ direction, namely, the direction of the electric field, the weak values should satisfy
\begin{eqnarray}
c\langle\hat{\sigma}_y\rangle_{\bf w}p_y+mc^2\langle\hat{\sigma}_z\rangle_{\bf w} = 0, \label{eq:cond_shift}
\end{eqnarray}
rather than $\langle\hat{\sigma}_y\rangle_{\bf w}=0$, where we have substituted $\hat{p}_y$ with $p_y$; the space part evolves into $\psi_{-p_x+q\varepsilon t,p_y}(x-c\langle\hat{\sigma}_x\rangle_{\bf w}t,y)$.
As a result, we obtain the same result as equation (\ref{eq:selective}):
the transition is selective i.e. from ($-E,-p_x,p_y$) to ($E,p_x,p_y$).
Then, the weak value of $\hat{\sigma}_x$ is given by
\begin{eqnarray}
\langle \hat{\sigma}_x\rangle_{\bf w} = \frac{\sqrt{p_x^2+p_y^2+m^2c^2}}{p_x} = \frac{E}{p_xc},  \label{eq:WV2m}
\end{eqnarray}
where we have made use of $E^2=p^2c^2+m^2c^4=(p_x^2+p_y^2)c^2 + m^2c^4$.

\begin{figure}
  \begin{center}
	 \includegraphics[scale=0.55]{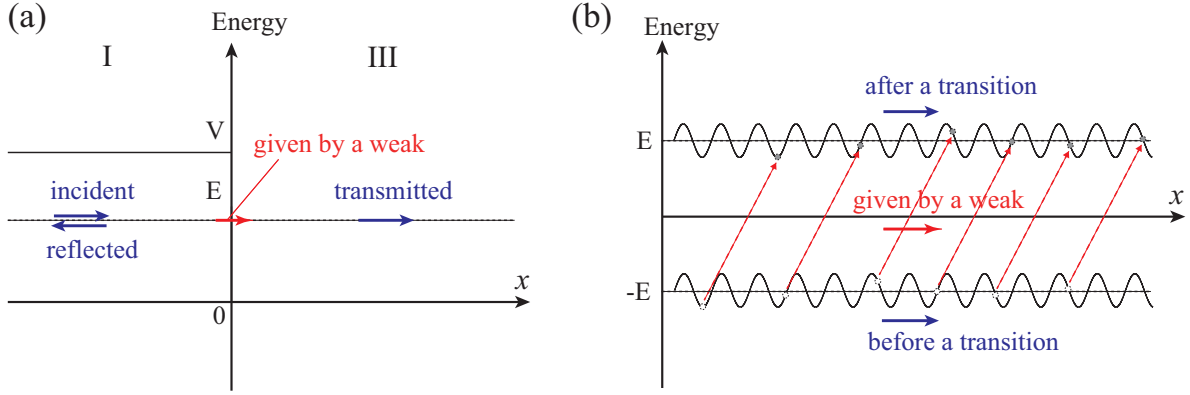}
  \end{center}
  \caption{The consistencies of the weak values in (a)the transmission and (b)the transition.
(a)An incident particle in I arrives at the step of the potential $V$ ($x=0$) and may transmit to III, at the moment of which the velocity is given by a weak value.
The average velocity of the steady flux of particles in I and III comes from generating the weak value of the group velocity at the step.
The consistency should be satisfied between these velocities.
(b)Unlike the transmission, a transition occurs all over the space with the coherence. Then, the process described by a weak value represents the transition between the two wave functions of the energy eigenstates, which are illustrated by the sinusoidal curves.
In this case, the weak value should be consistent with the average velocity of the waves. Then, the phase velocity (rather than the group velocity) will be appropriate for such purpose, since they contain the property of the waves like the coherence.
}
\label{fig:difference}
\end{figure}

On equation (\ref{eq:cond_shift}), we note that this case is different from the `transmission' in a supercritical step potential \cite{step}, in which the massive term $mc^2\langle\hat{\sigma}_k\rangle_{\bf w}$ is not eliminated.
In the `transmission,' a pair production takes place at the step, in which how to achieve the time (and the space) evolution is described by a weak value as shown in figure \ref{fig:difference} (a).
The wave function represents the steady flux of the particles, which are incident to the step and are transmitted or reflected at the step.
Then, with the transmission probability, the weak value of a group velocity should be consistent with the the group velocities in I and III.
On the other hand, in the `transition' shown in figure \ref{fig:difference} (b), a pair production may take place all over the space, whose probability is uniformly distributed.
What the weak value provides is the process of the transition between the two wave functions in a negative energy and a positive one, which represents the fields governing the behavior of a particle in a sense:
where such process takes place is not determined beforehand due to the coherence all over the space.
Unlike the quantum time evolution of one wave function in the `transmission,' the time evolution in the `transition' should conventionally represent how the electric field achieves the real work and the real impulse.
As a result, it is given by the merely shift without the additional shift i.e. the massive term $mc^2\langle\hat{\sigma}_z\rangle_{\bf w}$ as shown in equation (\ref{eq:cond_shift}).

The consistency of the average velocity like equation (\ref{eq:cons_WV2}) must be also on the two waves, which will be on the phase velocities rather than the group velocities.
Then, using the transition probability $T_{2+1}(p)$, the weak value of a group velocity, which exactly corresponds to the shift in $x$ direction i.e. the phase velocity, should satisfy,
\begin{eqnarray}
T_{2+1}(p)c\langle\hat{\sigma}_x\rangle_{\bf w} = \frac{E}{p}\frac{p_x}{\sqrt{p_x^2+p_y^2}}, \label{eq:cons_WV2m}
\end{eqnarray}
where r.h.s. is the (average) phase velocity in $x$ direction just before and after the transition.
In consequence, we can obtain the same transition probability as equation (\ref{eq:prob_2}).
Unlike the massless case, the average velocity $T_{2+1}(p)c\langle\hat{\sigma}_x\rangle_{\bf w}$ may exceed $c$ as a phase velocity, while it is also the requisite velocity for the electric field to achieve both the work and the impulse.
Due to the coherence all over the space, however, such superluminal velocity is not available and does not contradict the causality.
In the massless case \cite{gra_wv}, despite the coherence, we could treat plane waves as if particles effectively like in transmission, because the group velocity always accorded with the phase one.
In such description of de facto particles, we could expect to see how a strange weak value (superluminal velocity) was avoided
\footnote{
It is not always strange that a group velocity also exceeds $c$.
However, the appearance of such velocity would be noteworthy in dealing with plane waves only.
}.
Considering the massive case, we have uncovered the difference between the transmission and the transition on this occasion.
However, in both cases, it is common that a weak value determines the physics (the transmission/transition probability) by keeping the consistency of the velocity (particles/waves).

\begin{figure}
  \begin{center}
	 \includegraphics[scale=0.8]{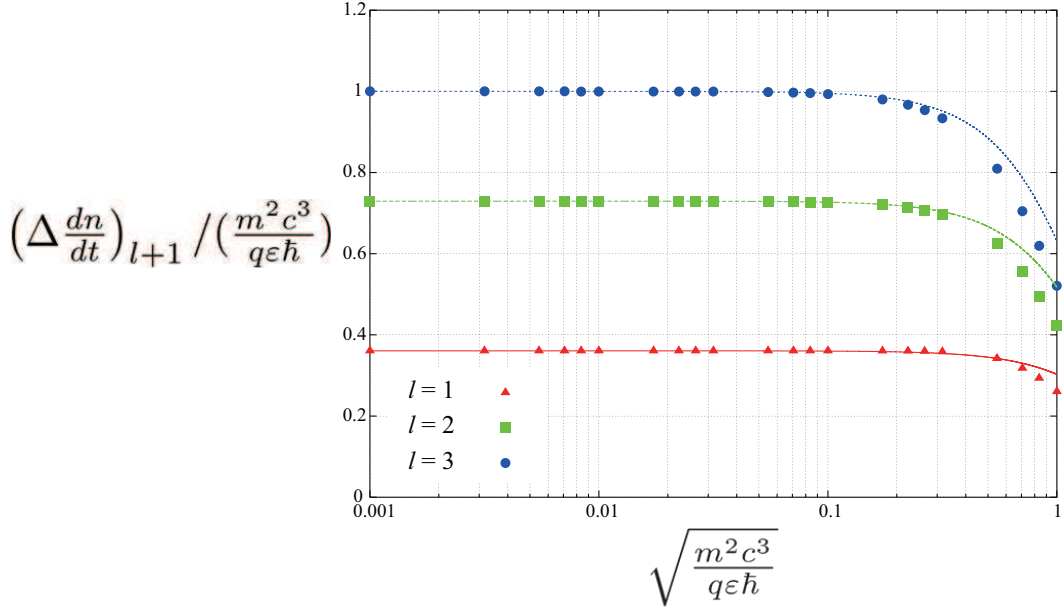}
  \end{center}
  \caption{The numerical results of the decreasing rate by the mass in each dimension.
The curved lines represent the exponential decrease shown in equation (\ref{eq:i_rate_m}), normalized by the numerical values at $\sqrt{\frac{m^2c^3}{q\varepsilon\hbar}}=0.001$.
In any dimension, when $\sqrt{\frac{m^2c^3}{q\varepsilon\hbar}}\ll 1$, the decreasing rate is proportional to $\frac{m^2c^3}{q\varepsilon\hbar}$ and agrees with equation (\ref{eq:i_rate_m}) within the order.
}
\label{fig:result}
\end{figure}

As in the case of the massless particles, quantum states to be counted should satisfy $\Delta t \le \delta t=\hbar/\Delta E$, which gives,
\begin{eqnarray}
p_x^2(p_x^2+p_y^2)+m^2c^2p_x^2 \le \frac{q^2\varepsilon^2\hbar^2}{16c^2}.   \label{eq:qc_2}
\end{eqnarray}
As a result, we can estimate the decreasing rate by the mass,
\begin{eqnarray}
\left(\Delta \frac{dn}{dt}\right)_{2+1} = \frac{\left(\frac{dn}{dt}\right)^{m=0}_{2+1}-\left(\frac{dn}{dt}\right)_{2+1}}{\left(\frac{dn}{dt}\right)^{m=0}_{2+1}},  \label{eq:rate_mass2}
\end{eqnarray}
with
\begin{eqnarray}
\left(\frac{dn}{dt}\right)_{2+1} &=& \frac{1}{(2\pi\hbar)^2}\int \!\!\! \int_{(\ref{eq:qc_2}) \ {\rm and} \ p_x>0}dp_xdp_y\frac{T_{2+1}(p)}{\Delta t} \\
&=& \frac{q\varepsilon}{2(2\pi\hbar)^2}\int \!\!\! \int_{(\ref{eq:qc_2}) \ {\rm and} \ p_x>0}dp_xdp_y\frac{p_x}{p_x^2+p_y^2}            \label{eq:rate2m}   \\
&=& \frac{(q\varepsilon)^{3/2}}{4\pi^2\hbar ^{3/2}c^{1/2}}\frac{1}{2}\int_{0}^{\sqrt{-2A+\sqrt{4A^2+1}}} ds \ {\rm Arctan}\sqrt{\frac{1}{s^4}-\frac{4A}{s^2}-1},
\end{eqnarray}
where $A$ is defined by $\frac{m^2c^3}{q\varepsilon\hbar}$.
If we substitute $m=0$ into the above equations, the same result of equation (\ref{eq:rate2_0}) can be derived:
the massive case contains the massless one, which is just obtained with $m=0$.
Figure \ref{fig:result} ($l=2$) shows the numerical results of equation (\ref{eq:rate_mass2}), which indicate that the decreasing rate by the mass in $\sqrt{\frac{m^2c^3}{q\varepsilon\hbar}}\ll 1$ is proportional to $\frac{m^2c^3}{q\varepsilon\hbar}$ and about corresponds to equation (\ref{eq:i_rate_m}).
Note that $\sqrt{\frac{m^2c^3}{q\varepsilon\hbar}}\ll 1$ $\longrightarrow$  $\frac{m^2c^3}{q\varepsilon\hbar}\ll 1$, in which the approximation of equation (\ref{eq:i_rate_m}) is valid.

According to equation (\ref{eq:qc_2}), $p_x$ is at most $\frac{1}{2}\sqrt{\frac{q\varepsilon\hbar}{c}}$.
Then, the applicable scope of our model, namely, $\sqrt{\frac{m^2c^3}{q\varepsilon\hbar}}\ll 1$ turns out to be
\begin{eqnarray}
mc^2\ll \sqrt{q\varepsilon\hbar c} < \frac{q\varepsilon\hbar}{2p_x}=\frac{\hbar}{\Delta t},    \label{eq:cond_gap}
\end{eqnarray}
which means that the mass energy is much smaller than the energy fluctuation, $\hbar/\Delta t$.
In the massive case, there is the mass energy gap between $-mc^2$ and $mc^2$;
the transition through the energy gap can be understood in the context of the Landau-Zener tunneling, which brings about the exponential decrease in the pair production rate as shown in equation (\ref{eq:i_rate}).
It is plausible that such exponential decrease is approximated to the first order of $\frac{m^2c^3}{q\varepsilon\hbar}$, when the tunneling gap can be regarded as small enough to satisfy (\ref{eq:cond_gap}), which is achieved in the high electric field $\varepsilon$.
In figure \ref{fig:result} ($l=2$), the agreement in $\sqrt{\frac{m^2c^3}{q\varepsilon\hbar}}\ll 1$ ensures that our model of the virtual transition gives us a new aspect to explain the exponential decreasing rate of the Landau-Zener tunneling in the high electric field as follows.
Mathematically the mass energy gap decreases the integral range in equation (\ref{eq:rate2m}) and, eventually, the number of quantum states to be counted, because the integrand in equation (\ref{eq:rate2m}) is same as the one in the massless case as shown in equation (\ref{eq:rate2}).
The decrease of the number depends on the electric field, which effectively determines the width of the tunnel (gap) caused by the mass energy: as the electric field becomes smaller, the width of the tunnel seems to be wider relatively.
The decreasing rate of the Landau-Zener tunneling in the high electric field can be explained with such decreasing number of quantum states contributing to the pair productions.

\section{The 3+1 dimensional case}
\label{sec:3+1}
Straightforwardly, we can apply the procedure mentioned in the previous section to the 3+1 dimensional case, in which the Hamiltonian is given by
\begin{eqnarray}
\hat{H}_{3+1}=c(\hat{\alpha}_x\hat{p}_x+\hat{\alpha}_y\hat{p}_y+\hat{\alpha}_z\hat{p}_z)+mc^2\hat{\beta}-q\varepsilon x,
\end{eqnarray}
with the uniform electric field $\varepsilon$ in $x$ direction and
\begin{eqnarray}
\hat{\alpha}_k=\left[ 
 \begin{array}{cc}
 \ & \hat{\sigma}_k \\
 \hat{\sigma}_k & \ \\
 \end{array} 
\right], \ \ \ 
\hat{\beta}=\left[ 
 \begin{array}{cc}
 \hat{1} & \ \\
 \ & -\hat{1} \\
 \end{array} 
\right],
\end{eqnarray}
where $\hat{1}$ denotes the $2\times 2$ identity matrix.
An eigenstate of the free Hamiltonian can also be described by a chirality and a space part:
a negative (positive) energy eigenstate with an energy $-E$ ($E$) and a momentum $\vec{p}=(p_x,p_y,p_z)$ can be assigned with the chirality as follows,
\begin{eqnarray}
|-E,p_x,p_y,p_x\rangle = a_-\left[
 \begin{array}{c}
 -\frac{p_zc}{E+mc^2} \\
 -\frac{(p_x+ip_y)c}{E+mc^2} \\
 1 \\
 0
 \end{array}
\right]
+b_-\left[
 \begin{array}{c}
 -\frac{(p_x-ip_y)c}{E+mc^2} \\
 \frac{p_zc}{E+mc^2} \\
 0 \\
 1
 \end{array}
\right] \\
|E,p_x,p_y,p_z\rangle = a_+\left[
 \begin{array}{c}
 1 \\
 0 \\
 \frac{p_zc}{E+mc^2} \\
 \frac{(p_x+ip_y)c}{E+mc^2} 
 \end{array}
\right]
+b_+\left[
 \begin{array}{c}
 0 \\
 1 \\
 \frac{(p_x-ip_y)c}{E+mc^2} \\
 -\frac{p_zc}{E+mc^2} 
 \end{array}
\right].
\end{eqnarray}
Note that a Dirac particle has a half spin in 3+1 dimension unlike 1+1 and 2+1 dimensions;
the arbitrary coefficients, $a_{\pm}$ and $b_{\pm}$, determine the direction of the spin in the non-relativistic term.
We also consider a pair production by the electric field, which is represented by the transition from a negative energy state $(-E,-p_x,p_y,p_z)$ to a positive one ($E',p_x',p_y,p_z$) with $E,E',p_x,p_x'$ $>0$, where the momentums in $y$ and $z$ are constants of motion.
Then, the space part $\psi_{-p_x,p_y,p_z}(x,y,z)$ evolves as follows,
\begin{eqnarray}
& &\langle E',p_x',p_y,p_z| e^{-\frac{i}{\hbar}\hat{H}_{3+1}t}|-E,-p_x,p_y,p_z\rangle \psi_{-p_x,p_y,p_z}(x,y,z)               \\
&\sim& \langle E',p_x',p_y,p_z|-E,-p_x,p_y,p_z\rangle e^{-\frac{i}{\hbar}c\langle\hat{\alpha}_x\rangle _{\bf w}\hat{p}_xt}e^{-\frac{i}{\hbar}c\langle\hat{\alpha}_y\rangle_{\bf w}\hat{p}_yt} \nonumber \\
& & \ \ \ \ \ \ \ \ \ \ \ \ \ \ e^{-\frac{i}{\hbar}c\langle\hat{\alpha}_z\rangle_{\bf w}\hat{p}_zt}e^{-\frac{i}{\hbar}mc^2\langle\hat{\beta}\rangle_{\bf w}t}e^{\frac{i}{\hbar}q\varepsilon x t}\psi_{-p_x,p_y,p_z}(x,y,z)  \ \ \ (t \ \sim \ 0) \\
&\sim& \langle E',p_x',p_y,p_z|-E,-p_x,p_y,p_z\rangle e^{-\frac{i}{\hbar}mc^2\langle\hat{\beta}\rangle_{\bf w}t} \nonumber \\
& & \ \ \ \psi_{-p_x+q\varepsilon xt,p_y,p_z}(x-c\langle\hat{\alpha}_x\rangle_{\bf w}t, y-c\langle\hat{\alpha}_y\rangle_{\bf w}t, z-c\langle\hat{\alpha}_z\rangle_{\bf w}t) \ \ \ (t \ \sim \ 0),
\end{eqnarray}
where $\langle\hat{\alpha}_k\rangle_{\bf w}$ and $\langle\hat{\beta}\rangle_{\bf w}$ are weak values defined as in equation (\ref{eq:WV});
$c\langle\hat{\alpha}_k\rangle_{\bf w}$ gives the velocity in $k$ direction.
From the condition of the shift in $x$ direction:
\begin{eqnarray}
c\langle\hat{\alpha}_y\rangle_{\bf w}p_y+c\langle\hat{\alpha}_z\rangle_{\bf w}p_z+mc^2\langle\hat{\beta}\rangle_{\bf w}=0,
\end{eqnarray}
we can derive the same result as equation (\ref{eq:selective}).
That is, the transition is selective to be $(-E,-p_x,p_y,p_z)$ $\rightarrow$ $(E,p_x,p_y,p_z)$.
This result can be found regardless of the coefficients, $a_{\pm}$ and $b_{\pm}$:
with weak values, we cannot determine the directions of the spins of pairs, which should be subjected to something else.
However, this is enough to estimate the production rate, because the two degrees of freedom by the half spin get involved with the rate just twice as shown in equation (\ref{eq:i_rate3}).  

With the weak value of $\hat{\alpha}_x$ given by
\begin{eqnarray}
\langle \hat{\alpha}_x\rangle_{\bf w} = \frac{E}{p_xc}\left(=\frac{\sqrt{p_x^2+p_y^2+p_z^2+m^2c^2}}{p_x}\right),  \label{eq:WV3m}
\end{eqnarray}
the consistency of the velocity in $x$ direction provides the transition probability $T_{3+1}(p)$ as follows,
\begin{eqnarray}
T_{3+1}(p)c\langle\hat{\alpha}_x\rangle_{\bf w} = \frac{E}{p}\frac{p_x}{\sqrt{p_x^2+p_y^2+p_z^2}}, \label{eq:cons_WV3m}
\end{eqnarray}
where r.h.s. represents the (average) phase velocity just before and after the transition in $x$ direction.
Then, we can obtain,
\begin{eqnarray}
T_{3+1}(p)=\frac{p_x^2}{p_x^2+p_y^2+p_z^2}.
\end{eqnarray}

As we mentioned in the previous section, quantum states can contribute to pair productions via virtual transitions, which satisfy $\Delta t\le \delta t=\hbar/\Delta E$:
\begin{eqnarray}
p_x^2(p_x^2+p_y^2+p_z^2)+m^2c^2p_x^2\le\frac{q^2\varepsilon^2\hbar^2}{16c^2}.   \label{eq:qc_3}
\end{eqnarray}
As a result, we can estimate the decreasing rate by the mass,
\begin{eqnarray}
\left(\Delta \frac{dn}{dt}\right)_{3+1} = \frac{\left(\frac{dn}{dt}\right)^{m=0}_{3+1}-\left(\frac{dn}{dt}\right)_{3+1}}{\left(\frac{dn}{dt}\right)^{m=0}_{3+1}},  \label{eq:rate_mass3}
\end{eqnarray}
with
\begin{eqnarray}
\left(\frac{dn}{dt}\right)_{3+1} &=& 2\frac{1}{(2\pi\hbar)^3}\int \!\!\! \int \!\!\! \int_{(\ref{eq:qc_3}) \ {\rm and} \ p_x>0}dp_xdp_ydp_z\frac{T_{3+1}(p)}{\Delta t}    \label{eq:rate3m}  \\
&=& \frac{q\varepsilon}{(2\pi\hbar)^3}\int \!\!\! \int \!\!\! \int_{(\ref{eq:qc_3}) \ {\rm and} \ p_x>0}dp_xdp_ydp_z\frac{p_x}{p_x^2+p_y^2+p_z^2} \\
&=& \frac{(q\varepsilon )^2}{4\pi^3\hbar^2c} \frac{\pi}{8}\int_0^{\sqrt{-2A+\sqrt{4A^2+1}}}ds \ s \ {\rm ln}\left(\frac{1}{s^4}-\frac{4A}{s^2}\right), \label{eq:rate3m_re}
\end{eqnarray}
where $A\equiv\frac{m^2c^3}{q\varepsilon\hbar}$.
Note that we have counted the quantum states twice due to the spin.
Substituting $m=0$ into equation (\ref{eq:rate3m_re}), we can obtain the massless pair production rate as follows,
\begin{eqnarray}
\left(\frac{dn}{dt}\right)_{3+1}^{m=0}=\frac{\pi}{8}\frac{(q\varepsilon )^2}{4\pi^3\hbar^2c} \sim 0.39\frac{(q\varepsilon )^2}{4\pi^3\hbar^2c},  \label{eq:rate3_0}
\end{eqnarray}
which approximately corresponds to equation (\ref{eq:i_rate3}) with $m=0$. 
We also show the numerical results of equation (\ref{eq:rate_mass3}) in figure \ref{fig:result} ($l=3$).
Clearly, we can confirm $\left(\Delta\frac{dn}{dt}\right)_{3+1} \propto \frac{m^2c^3}{q\varepsilon\hbar}$ in $\sqrt{\frac{m^2c^3}{q\varepsilon\hbar}}\ll 1$, which agrees with equation (\ref{eq:i_rate_m}).

\section{The 1+1 dimensional case}
\label{sec:1+1}
In 1+1 dimension, we will soon notice that we need to contrive the particular estimation due to the singularity of the massless pair production.
We also begin with the Hamiltonian of a massive Dirac particle as follows,
\begin{eqnarray}
\hat{H}_{1+1}=c\hat{\sigma}_x\hat{p}_x+mc^2\hat{\sigma}_z-q\varepsilon x
\end{eqnarray}
in the uniform electric field $\varepsilon$ in $x$ direction.
As we have seen, a negative (positive) energy eigenstate of the free Hamiltonian is specified by the chirality $|-E,p_x\rangle$ ($|E,p_x\rangle$) as follows,
\begin{eqnarray}
|\pm E,p_x\rangle = \frac{1}{\sqrt{2E(E\mp mc^2)}}
\left[ 
\begin{array}{c}
 p_xc \\
 \pm E-mc^2 \\
 \end{array}
\right],   \label{eq:eigen1m}
\end{eqnarray}
with an energy $-E<0$ ($E>0$) and a momentum $p_x$
\footnote{
The choice of the Pauli matrices in the Hamiltonian is different from the one in \cite{step} and the chiralities differ too.
However, our results are not affected by the choices, because this is just on the representation.
}.
In the transition from ($-E,-p_x$) to ($E',p'_x$) with $E,E',p_x,p'_x$ $>0$ to represent a pair production, the Dirac particle moves along with the time evolution of the space part as follows,
\begin{eqnarray}
& &\langle E',p_x'| e^{-\frac{i}{\hbar}\hat{H}_{1+1}t}|-E,-p_x\rangle \psi_{-p_x}(x)               \\
&\sim& \langle E',p_x'|-E,-p_x\rangle e^{-\frac{i}{\hbar}c\langle\hat{\sigma}_x\rangle _{\bf w}\hat{p}_xt}
e^{-\frac{i}{\hbar}mc^2\langle\hat{\sigma}_z\rangle_{\bf w}t}
e^{\frac{i}{\hbar}q\varepsilon x t}\psi_{-p_x}(x)  \ \ \ (t \ \sim \ 0) \\
&\sim& \langle E',p_x'|-E,-p_x\rangle e^{-\frac{i}{\hbar}mc^2\langle\hat{\sigma}_z\rangle_{\bf w}t} \psi_{-p_x+q\varepsilon t}(x-c\langle\hat{\sigma}_x\rangle_{\bf w}t) \ \ \ (t \ \sim \ 0).
\end{eqnarray}
From the condition of the shift without an additional phase:
\begin{eqnarray}
mc^2\langle\hat{\sigma}_z\rangle _{\bf w} = 0, \label{eq:cond_shift1}
\end{eqnarray}
we can obtain the same result as equation (\ref{eq:selective}).
That is, only the transition from ($-E,-p_x$) to ($E,p_x$) is allowed.

The transition probability $T_{1+1}(p)$ can be derived from the consistency on the average velocities as follows,
\begin{eqnarray}
T_{1+1}(p)c\langle\hat{\sigma}_x\rangle_{\bf w} = \frac{E}{p_x}    \label{eq:cons_WV1}  \\ 
T_{1+1}(p) = 1,   \label{eq:trans1}
\end{eqnarray}
where, in equation (\ref{eq:cons_WV1}), $c\langle\hat{\sigma}_x\rangle_{\bf w}=E/p_x$ provides the velocity in the transition and r.h.s. is the (average) phase velocity immediately before and after the transition.

From $\Delta t\le\delta t=\hbar/\Delta E$, we can derive
\begin{eqnarray}
p_x^4+m^2c^2p_x^2\le\frac{q^2\varepsilon^2\hbar^2}{16c^2},   \label{eq:qc_1}
\end{eqnarray}
which represents the condition for quantum states to participate in pair productions.
Then, the decreasing rate by the mass can be calculated as follows,
\begin{eqnarray}
\left(\Delta\frac{dn}{dt}\right)_{1+1} = \frac{\left(\frac{dn}{dt}\right)^{m=0}_{1+1}-\left(\frac{dn}{dt}\right)_{1+1}}{\left(\frac{dn}{dt}\right)^{m=0}_{1+1}},  \label{eq:rate_mass1}
\end{eqnarray}
where the rate is given by
\begin{eqnarray}
\left(\frac{dn}{dt}\right)_{1+1} &=& \frac{1}{2\pi\hbar}\int_{(\ref{eq:qc_1}) \ {\rm and} \ p_x>0}dp_x\frac{T_{1+1}(p)}{\Delta t} \\
&=& \frac{q\varepsilon}{2(2\pi\hbar)}\int_{(\ref{eq:qc_1}) \ {\rm and} \ p_x>0}dp_x\frac{1}{p_x},   \\
&=& \frac{q\varepsilon}{2\pi\hbar}\frac{1}{2}\int_0^{\sqrt{-2A+\sqrt{4A^2+1}}}ds\frac{1}{s},    \label{eq:rate1_mass}
\end{eqnarray}
with $A\equiv\frac{m^2c^3}{q\varepsilon\hbar}$.
In particular, the massless pair production rate results in
\begin{eqnarray}
\left(\frac{dn}{dt}\right)_{1+1}^{m=0}=\frac{q\varepsilon}{2\pi\hbar}\frac{1}{2}\int_0^1 ds \ \frac{1}{s}.  \label{eq:rate1_0}
\end{eqnarray}
Clearly, (\ref{eq:rate1_mass}) and (\ref{eq:rate1_0}) diverge, while the difference between them:
\begin{eqnarray}
\left(\frac{dn}{dt}\right)_{1+1}^{m=0}-\left(\frac{dn}{dt}\right)_{1+1}=-\frac{q\varepsilon}{2\pi\hbar}\frac{1}{2}\int_1^{\sqrt{-2A+\sqrt{4A^2+1}}}ds \ \frac{1}{s},   \label{eq:diff1}
\end{eqnarray}
does not, which is the numerator of r.h.s. in equation (\ref{eq:rate_mass1}).
As we mentioned at the end of section \ref{sec:2+1}, the mass energy gap results in the decrease of quantum states contributing to the pair productions;
because such decrease itself explains the first order of the exponential decreasing rate of the Landau-Zener tunneling, it is plausible that the difference between the massless and massive pair production rates, namely, equation (\ref{eq:diff1}) settles in a certain value.
Then, what we need is how to determine the massless pair production rate which is also the denominator of r.h.s. in equation (\ref{eq:rate_mass1}).
In fact, this anomaly comes from the non-applicability of the condition given by equation (\ref{eq:cond_shift1}) in the massless case:
unlike the other cases, such condition does not appear in the massless, 1+1 dimensional case, because the Hamiltonian given by $\hat{H}=c\hat{\sigma}_x\hat{p}_x$ always brings about the shift in $x$ direction without an additional phase.
As a result, the transition is not selective in this case.

We reconsider the massless, 1+1 dimensional case ad hoc.
As shown in equation (\ref{eq:eigen1m}), the chiralities of the positive and negative energy eigenstates with the momentum $p_x$ are given by
\begin{eqnarray}
|\pm E,p_x\rangle = \frac{1}{\sqrt 2}
\left[
\begin{array}{c}
 p_x/|p_x| \\
 \pm 1 \\
 \end{array}
\right],
\end{eqnarray}
because of $E^2=p_x^2c^2$ and $m=0$.
Then, in any transition from a negative energy state ($-E,-p_x$) to a positive one ($E',p_x'$) where we have assumed $E,E',p_x,p'_x$ $>0$, the weak value is given by
\begin{eqnarray}
\langle\hat{\sigma}_x\rangle_{\bf w}=1,
\end{eqnarray}
which means that the velocity by the weak value is $c$.
Because the velocity of a massless Dirac particle is always $c$ like a photon primarily, the velocities just before and after the transition are also $c$.
Then, due to the consistency of the velocities, the transition probability $T_{1+1}(p)$ is always $1$, which is the same as equation (\ref{eq:trans1}), in any transition.

Because the post-selection is not uniquely determined for the pre-selection $|-E,-p_x\rangle$, we rely on the fluctuations by the uncertainty relations to estimate the average pair production rate $T_{1+1}(p)/\overline{\Delta t}(p_x)=1/\overline{\Delta t}(p_x)$ as follows.
According to the uncertainty relation, the transition energy $\Delta E$ can be allowed within the time $\delta t$ which satisfies
\begin{eqnarray}
\Delta E \delta t = \hbar.
\end{eqnarray}
In the transition from ($-E,-p_x$) to ($E',p'_x$), it gives $\Delta E$ $=$ $E+E'$ $=$ $c(p_x+p'_x)$.
We assume that the transition with the needed time $\Delta t(p_x,p'_x)$ satisfying $\Delta t(p_x,p'_x)\leq\delta t$ occurs with equally probability.
Since $\Delta t$ is given by $(p_x+p'_x)/(q\varepsilon)$ due to the relationship between the impulse and the momentum change: $q\varepsilon\Delta t=p_x+p'_x$,
we can derive $p'_x$ $\leq$ $\sqrt{q\varepsilon\hbar/c}-p_x$ from $\Delta t(p_x,p'_x) \leq \delta t$.
Then, the average transition time can be estimated as follow,
\begin{eqnarray}
\overline{\Delta t}(p_x) &=& \frac{\int_0^{\sqrt{q\varepsilon\hbar/c}-p_x}dp_x'\Delta t(p_x,p'_x)}{\int_0^{\sqrt{q\varepsilon\hbar/c}-p_x}dp_x'}  \nonumber \\
&=& \frac{1}{2q\varepsilon}\left(\sqrt{\frac{q\varepsilon\hbar}{c}}+p_x\right).
\end{eqnarray}
As a result, the average pair production rate can be given as follows,
\begin{eqnarray}
\left(\overline{\frac{dn}{dt}}\right)_{1+1}^{m=0} &=& \frac{1}{2\pi\hbar}\int_0^{\sqrt{\frac{q\varepsilon\hbar}{c}}}dp_x\frac{T_{1+1}(p)}{\overline{\Delta t}(p_x)}  \\
&=& 2{\rm ln}2\frac{q\varepsilon}{2\pi\hbar} \sim 1.39 \frac{q\varepsilon}{2\pi\hbar},
\end{eqnarray}
where the upper bound of the integration has been taken as $\sqrt{q\varepsilon\hbar/c}$, outside which we cannot define $\overline{\Delta t}(p_x)$.
This result approximately agrees with the massless pair production rate as shown in equation (\ref{eq:i_rate}) with $l=1$ and $m=0$.
Using this massless rate as the normalization, the decreasing rate by the mass can be re-defined as follows,
\begin{eqnarray}
\left(\Delta\frac{dn}{dt}\right)_{1+1} = \frac{\left(\frac{dn}{dt}\right)^{m=0}_{1+1}-\left(\frac{dn}{dt}\right)_{1+1}}{\left(\overline{\frac{dn}{dt}}\right)^{m=0}_{1+1}},  \label{eq:rate_mass1c}
\end{eqnarray}
where the numerator takes a finite value as already stated.
The numerical results of equation (\ref{eq:rate_mass1c}) are shown in figure \ref{fig:result} ($l=1$), and we can find that the decreasing rate by the mass is proportional to $\frac{m^2c^3}{q\varepsilon\hbar}$ in $\sqrt{\frac{m^2c^3}{q\varepsilon\hbar}}\ll 1$ as in the other dimensional cases.

\section{Conclusion}
\label{sec:conclusion}
A weak value appears as the velocity of a Dirac particle in the short time approximation of the time evolution.
As an example, we have discussed the Schwinger mechanism in various dimensions:
the pair production, which had its root in a virtual transition by the uncertainty relations, was well described by a weak value.
To estimate the pair production rates, we considered how a weak value was consistent with the velocities before and after the process, by which the transition probabilities were obtained.
Our model is available for not only the massless pair productions but also the massive pair productions in the high electric field.
In this case, we can regard the mass energy gap as so small that the exponential decrease of the Landau-Zener tunneling probability can be approximated to the first order of $\frac{mc^3}{q\varepsilon\hbar}$;
the decreasing rate comes from the decrease of the number of the quantum states participating in the pair productions due to the forbidden mass energy gap.
As a result, in the high electric field, the decreasing rate of the Landau-Zener tunneling is associated with energy fluctuations via the weak values.
Considering the consistency of the weak values in the massive particles, we have also clarified the difference between the transmission and the transition.
In these cases, once a weak value is accepted as a value of a physical quantity and is interfaced with the other values, we succeed in understanding the quantum phenomena quantitatively, which also gives us simple descriptions to grasp them qualitatively.

\section*{Acknowledgements}
This work was supported by JSPS Grant-in-Aid for Scientific Research(A) 25247068.

\end{document}